\documentclass[conference]{IEEEtran}

\usepackage{cite}
\usepackage[pdftex]{graphicx}

\usepackage{tikz}
\usepackage{amsmath}
\usepackage{multicol}
\usepackage{multirow}
\usepackage{fontawesome}
\usepackage{array}
\usepackage{mathtools}
\usepackage{colortbl}
\usepackage{xspace}
\usepackage{kmath}
\usepackage{booktabs}
\usepackage[hidelinks]{hyperref}

\PassOptionsToPackage{table,dvipsnames,x11names}{xcolor}

\usepackage{colortbl}
\usepackage{pgfplots}
\pgfplotsset{compat=1.17}  % Ensure compatibility for pgfplots

\usepackage{xcolor}

\definecolor{mygray}{HTML}{ededed}

\usepackage{longtable}

\newcommand{\verttext}[1]{
  \parbox[t]{2mm}{\rotatebox[origin=lB]{60}{#1}}
}
\newcommand{\yes}{$\medbullet$\xspace}
\newcommand{\no}{$-$}
\newcommand{\leftrule}[1]{
    \multicolumn{1}{|c}{#1}
}

\begin{document}
\title{Evaluating Privacy Measures in Healthcare Apps Predominantly Used by Older Adults}

% % author names and affiliations
% % use a multiple column layout for up to three different
% % affiliations
% \author{\IEEEauthorblockN{Anonymous}}

\author{\IEEEauthorblockN{Suleiman Saka}
\IEEEauthorblockA{\textit{Department of Computer Science} \\
\textit{University of Denver}\\
Denver, US \\
suleiman.saka@du.edu}
\and
\IEEEauthorblockN{Sanchari Das}
\IEEEauthorblockA{\textit{Department of Computer Science} \\
\textit{University of Denver}\\
Denver, US\\
sanchari.das@du.edu}
}
% \author{
%     \IEEEauthorblockN{Suleiman Saka\textsuperscript{1}, Sanchari Das\textsuperscript{1,2}} 
%     \IEEEauthorblockA{
%         \textsuperscript{1}\textit{Department of Computer Science, University of Denver, USA}\\
%         \underline{suleiman.saka@du.edu} , \underline{sanchari.das@du.edu}
%     }
%     \IEEEauthorblockA{
%         \textsuperscript{2}\textit{Information Sciences and Technology Department, George Mason University, USA}\\
%         \underline{{sdas35}@gmu.edu}
%     }
% }

% \IEEEoverridecommandlockouts
% \makeatletter\def\@IEEEpubidpullup{6.5\baselineskip}\makeatother
\IEEEpubid{\parbox{\columnwidth}{
    BuildSEC'24 Building a Secure \& Empowered Cyberspace\\
    19-21 December 2024, New Delhi, India\\
    % ISBN 1-891562-93-2\\
    % https://dx.doi.org/10.14722/ndss.2025.23xxx\\
    https://www.buildsec.org/\\
}
\hspace{\columnsep}\makebox[\columnwidth]{}}

\maketitle

\begin{abstract}
The widespread adoption of telehealth systems has led to a significant increase in the use of healthcare apps among older adults, but this rapid growth has also heightened concerns about the privacy of their health information. While HIPAA in the US and GDPR in the EU establish essential privacy protections for health information, limited research exists on the effectiveness of healthcare app privacy policies, particularly those used predominantly by older adults. To address this, we evaluated $28$ healthcare apps across multiple dimensions, including regulatory compliance, data handling practices, and privacy-focused usability. To do this, we created a Privacy Risk Assessment Framework (PRAF) and used it to evaluate the privacy risks associated with these healthcare apps designed for older adults. Our analysis revealed significant gaps in compliance with privacy standards to such, only $25\%$ of apps explicitly state compliance with HIPAA, and only $18\%$ mention GDPR. Surprisingly, $79\%$ of these applications lack breach protocols, putting older adults at risk in the event of a data breach.
\end{abstract}

\begin{IEEEkeywords}
Privacy, Older Adults, Healthcare Apps 
\end{IEEEkeywords}

\section{Introduction}
The use of healthcare apps has soared exponentially in recent years~\cite{tazi2024evaluating}, with over $325,000$ health-related mobile apps available as of $2022$~\cite{Francoise_AppHealt_2022,surani2022understanding,tazi2024sok}. 
The global mobile health market is currently valued at $\$181$ billion and is anticipated to surpass $\$300$ billion by $2025$~\cite{TotalMHealth}. An important demographic that stands to benefit from this technology are older adults. As this demographic grows~\cite{us_census}, more of them are expected to use digital health applications for various needs~\cite{birkhoff_challenges_2020, Albrecht2018ChancenUR, Fan2020AnEE,das2022privacy}. However, the rapid development and adoption of such technology raise significant concerns about the privacy of personal health information~\cite{saka2023safeguarding}. Issues such as inadequate compliance with privacy policies~\cite{Parker2017AHA, Dehling2015ExploringTF,tazi2022sok,tazi2024exploring}, and improper data handling highlight the need to assess the efficacy and compliance of app privacy policies to ensure adequate protection of older adults' health information~\cite{Papageorgiou2018SecurityAP, Kapoor2022SilverSO, Adhikari2014SecurityAP,tazi2024we}, yet remain understudied. 
To address this, we evaluated the privacy policies of $28$ healthcare apps predominantly used by older adults by creating the Privacy Risk Assessment Framework (PRAF). This framework was established on regulations such as the Health Insurance Portability and Accountability Act (HIPAA)~\cite{Saha2023HIPAACheckerTC, hhsSummaryHIPAA,tazi2023privacy,tazi2023sok} and General Data Protection Regulation (GDPR)~\cite{gdpr_2016,surani2023security,adhikari2023evolution,tazi2024improving}, aiming to identify any concern in the privacy policies that could impact older adults. Our findings reveal significant variability in the comprehensibility and transparency of privacy policies and key privacy principles such as data minimization, retention period, and breach protocol were unevenly implemented across the apps. Through our analysis, we aimed to answer the following research questions:
\begin{itemize}
\item \textit{RQ1: How comprehensible and transparent are healthcare app privacy policies for older adult users?}
\item \textit{RQ2: What key principles or protections are missing from app policies when compared to privacy regulations?}
\item \textit{RQ3: What key trends emerge from the multi-dimensional privacy risk profiles of healthcare apps for older adults?}

\end{itemize}

\section{Background}
The growing reliance on healthcare applications and telehealth systems underscores the critical need for robust privacy measures for personal health information~\cite{Kharrazi2012MobilePH,monroe2021location}. Despite the establishment of regulations such as HIPAA~\cite{hhsSummaryHIPAA, Saha2023HIPAACheckerTC} in the United States and the GDPR in the European Union~\cite{gdpr_2016}, which set forth comprehensive protections for personal health information, there remains a notable gap in research concerning the effective compliance of these measures within healthcare apps targeted at older adults~\cite{zubaydi_mhealth_2015, Hordern_data_202, Muchagata_GDPR_2018}. Existing studies reveal a concerning trend: healthcare apps, especially those aimed at aiding older adults, often lack clear compliance and effective privacy policies~\cite{Papageorgiou2018SecurityAP}. Armontrout et al. reviewed the regulatory framework for mobile mental health apps and discovered that the majority of these apps were not subject to FDA regulation, emphasizing the need for more precise guidelines and standards for app developers~\cite{Armontrout_mhealth_2018}. Similarly, Rosenfeld et al. found that most health apps designed for dementia patients fail to incorporate any form of privacy policy, and those that do often present them in a manner that is difficult to comprehend~\cite{ROSENFELD_dementia_2017}. This observation is echoed by Blenner et al.~\cite{Blenner_Diabetes_2016} and Huckvale et al.~\cite{Huckvale_mental_2019}, who pointed out that some apps, including those for managing diabetes and mental health conditions, fall short of transparently communicating their data-sharing practices~\cite{Sampat_datashare, Grundy_dat_sharing_2019}.

Mulder compared app developers' marketing narratives with the content of their privacy policies, uncovering misalignment between promotional claims and actual policy~\cite{Trix_HealthApp_2019}. This discrepancy underscores the urgent need for privacy policies to conform with regulatory requirements, particularly within the healthcare domain. Our study expands on this work by examining HIPAA and GDPR compliance across a range of healthcare apps for older adults. 
% Alkhatib et al. advocate for a comprehensive approach to privacy in the deployment of Internet of Things (IoT) for older adults, emphasizing the importance of safeguarding confidentiality and regulating data usage~\cite{Alkhatib_privacy_2018}.
Birkhoff et al. investigated the unique challenges in mobile health app development, advocating for interprofessional strategies to address privacy and security concerns effectively~\cite{birkhoff_challenges_2020}. Similarly, Pywell et al. investigated the barriers to older adults' uptake of mobile-based mental health solutions, including concerns about data privacy and security, calling for the implementation of robust privacy policies as crucial steps toward protecting personal data~\cite{pywell_barriers_2020}. 
Despite these studies providing insightful information into various aspects of digital health privacy, the focus on healthcare apps specifically designed for older adults remains relatively underexplored. Our study aims to bridge the gap between healthcare app privacy studies and the unique needs of older adults by providing an analysis that reviews regulatory compliance, addresses age-specific concerns, and evaluates the implementation of key privacy concepts.
%-------------------------------------------------------------------------------
\section{Methodology}
%-------------------------------------------------------------------------------

\subsection{Selection of Healthcare Apps} 
Our list of healthcare apps for older adults was compiled through app stores because after surfing the Internet for a comprehensive list of healthcare apps specific to older adults we weren’t able to get a database that already has those healthcare apps. To collect the desired healthcare apps for our study we set inclusion criteria that include apps that are predominantly designed for older adults (age $65$ and above) and offer health-related functionalities (telehealth, medication tracking/reminder, vital sign sensors, etc.). 
% To obtain the desired list of healthcare apps for older adults and get access to their privacy policies, 
We defined our search strategy using keywords like \lq\lq Older Adults\rq\rq~ OR \lq\lq Seniors\rq\rq~ OR \lq\lq Elderly\rq\rq~ OR \lq\lq Geriatric\rq\rq~ in combination with \lq\lq Health\rq\rq~, OR \lq\lq Care\rq\rq~OR \lq\lq Healthcare\rq\rq~OR \lq\lq Telehealth\rq\rq~. We used our keywords in the two most popular app stores (Apple Store and Google Play Store) which resulted in $53$ apps.

\subsection{App Screening and Anonymization}
The collected apps were screened to eliminate duplicates, non-relevant apps and those that lack healthcare-related functions for older adults. In our preliminary screening, we narrowed down our collection from $53$ to $28$ healthcare apps. Another notable point about our final selection of 28 healthcare apps is that we ensured each app explicitly stated in its app store description that it is specifically designed for older adults.
% This final set of $28$ healthcare apps used by older adults was explored further to retrieve the most recent versions of app privacy policies for our analysis. 
A codebook was created to summarize the key information and also to extract the privacy policies' weblink.
To maintain ethical research practices and protect the reputation of the healthcare apps, we decided to use pseudonyms (A1, A2, A3,..., A28) when referencing them in our study. This approach is in line with methods used in related studies~\cite{ROSENFELD_dementia_2017, Elish_2020_ComRisk, denipitiyage_2024_mobileapp}, which use pseudonyms to protect companies from potential harm while allowing for a transparent discussion of the findings. 
% we use pseudonyms to help the industry improve its privacy practices without unfairly targeting specific businesses. 
We have already communicated our findings directly to the app developers, allowing them to fix any issues we uncover. 

\subsection{Privacy Risk Assessment Framework (PRAF)}
The Privacy Risk Assessment Framework was created to evaluate the privacy risks associated with healthcare apps designed for older adults. The framework is based on reviewing the privacy policy literature, existing data protection regulations such as HIPAA and GDPR, and considering the unique needs of older adults. PRAF is made up of five key elements, each chosen for its crucial role in protecting older adults' privacy. 
Due to the importance of regulatory compliance with established privacy standards, which are fundamental to data protection, we selected Regulatory Compliance as a key element. Given the highly confidential nature of health information and the increased vulnerability of older adults to data breaches~\cite{Nyteisha_privacy_2022, Ari_privacy_2022}, data security was selected. This element assesses the apps' security measures, including data encryption, access control, and breach protocol. To address the cognitive and physical limitations that may impair older adults' capacity to understand and manage privacy settings~\cite{Kuerbis_2017_OlderAA, Beach_privacy_2009}, usability and accessibility were included.  It assesses features like readability, language clarity, and the presence of accessibility features. Data minimization and Retention were essential to reduce privacy risks resulting from unnecessary data retention because this element measures the amount of data collection and storage practices. Third-Party Data sharing was added to measure the degree of control and transparency in data sharing with external entities. 
% which is a major policy challenge that arises from an interconnected health app environment.

The scoring system for each element represents their importance to the overall privacy risk assessment. 
The scoring varies among the five elements because some were assessed based on their presence or absence, while others were evaluated according to the level of adherence to specific standards.
Regulatory compliance is scored on a scale of 1-4, Data security is scored 1-2 for each criterion, usability and accessibility are scored on a scale of readability of 1-6, and clarity and accessibility on a scale of 1-2 each. Data minimization, retention and third-party data sharing were each scored on a scale of 1-2. A score of 1 indicates non-implementation, while a score of 2 indicates implementation.  The overall privacy risk score for each app is achieved by summing the scores from each element.
We used the PRAF to assess the $28$ healthcare apps based on each app's privacy policy. 
% This was done by manual review of policies and automated readability measures. The scores obtained allow us to have an overview of the privacy risks on each app and allow for comparisons among different apps as well as identify key trends and gaps.

% \begin{figure}[htbp!]
% \centering
% \includegraphics[width=1\linewidth]{Healthcare Apps/BuildSec_24/PRAF.drawio (1).pdf}
% \caption{Privacy Risk Assessment Framework (PRAF)}
% \label{PRAF}
% \end{figure}

\subsection{Privacy Policy Extraction}
To carry out a comprehensive analysis of healthcare apps targeted at older adults, we extracted the privacy policies from the collected apps. First, we collected privacy policy URLs from each app, and we created a Python script to automate the scraping of privacy policy content from their respective websites. 
% Because HIPAA, GDPR, and other data protection regulations are essential elements of the PRAF's regulatory compliance assessment, we gave priority to content related to these regulations during the extraction process. 
To ensure the accuracy of the data, we manually reviewed each privacy policy to validate the automated process and capture any information that might have been missed. We employed both methods, acknowledging that automated techniques may occasionally ignore specific aspects so we supplemented with a manual assessment to check the collected data and capture any nuances that the automated process might have missed.
The HIPAA privacy rule was established to safeguard the confidentiality and security of health records~\cite{hhsSummaryHIPAA}. However, its coverage is limited to Protected Health Information (PHI) managed by three entities: health plans, healthcare providers, and healthcare clearinghouses~\cite{nihHIPAAPrivacy}. Consequently, it does not extend to a broader spectrum of situations, leaving personal data held by other uncovered entities unregulated. 
Although just these three entities are covered by the HIPAA privacy rules, we included all collected apps in our research. This addition guarantees a thorough evaluation because it can be difficult to identify which particular apps are exempted. Furthermore, the GDPR has no such exemptions and requires extensive data protection procedures for all kinds of entities.
%-------------------------------------------------------------------------------
\section{Results}
%-------------------------------------------------------------------------------
Our analysis of the $28$ healthcare apps for older adults identified significant variations regarding privacy practices and regulatory compliance. 
% These include apps that provide functionality to track sleep, monitor physical activity, set medication reminders and access telehealth services.
Some of these apps did not explicitly mention HIPAA and GDPR but they had some mentions of other data regulations like the European Economic Area (EEA) rights, California Consumer Privacy Act (CCPA), and other international privacy regulations. Using the Privacy Risk Assessment Framework (PRAF) helped us to discover some key trends and gaps.
Among the $28$ apps, $7$ apps included privacy policies referencing HIPAA, $5$ apps mentioned GDPR, $12$ apps cited other data protection acts, and $4$ apps made no references to any privacy regulations. Additionally, one app, $A24$, had an inaccessible privacy policy. It is important to note that while some apps claim to adhere to privacy regulations, the actual effectiveness and compliance with these regulations cannot be guaranteed. Therefore, it is crucial to examine the details of each app's privacy policy. Table~\ref{tab2} shows the apps, usage, regulations, key principles, limitations, readability and privacy risk scores.

\begin{table*}[hbt!]
\caption{Healthcare App Privacy Policy Analysis, Readability using the SMOG Index and Privacy Risk Scores. Legend: Yes = \yes \textbar Partial = $\circ$  \textbar No = \no \textbar SD = Slightly Difficult, SWD = Somewhat Difficult, FD = Fairly Difficult, D = Difficult, VD = Very Difficult, P = Professional}
  \centering
  \setlength{\tabcolsep}{2.5pt}
      \footnotesize
\begin{tabular}{ l    
                l           
                % l
                % l
                % l
                *{3}{c}     
                *{6}{c}     
                *{4}{c}    
                *{2}{c}
                *{6}{c}
               }
               &
               &
               % &
               % &
               % &
               
\multicolumn{3}{c}{{\textbf{Regulation}}} &
\multicolumn{6}{c}{{\textbf{Key principles}}} &
\multicolumn{4}{c}{{\textbf{Limitations/Gaps}}} &
\multicolumn{2}{c}{{\textbf{Readability}}} &
\multicolumn{6}{c}{{\textbf{PRAF Scores}}} \\ 
 
     \textbf{Apps}  &  
     \textbf{Use} &
    \verttext{HIPAA}&
    \verttext{GDPR} &
     \verttext{Others} & 
 
    % Key principle
    \verttext{Data minimization} &
    \verttext{Data Encryption} &
    \verttext{Access controls} &
    \verttext{Consent requirements} &
    \verttext{Retention time} &
    \verttext{Breach protocol} &
    
    % Gaps
    \verttext{Ambiguous language} &
    \verttext{Vague commitments} &
    \verttext{Accessibility accomms} &
    \verttext{3rd party sharing} &

    % READIBILITY
        \verttext{SMOG} &
    \verttext{Level} &  

    %Privacy risk score
    \verttext{Regulatory Compliance} &
    \verttext{Data Security} &
    \verttext{Usability/Accessibility} &
    \verttext{Minimization/Retention} &
    \verttext{3rd party sharing} &
    \verttext{Overall Risk} \\    
    \midrule
       
    % %
    %%%%%%%%%%%%%%%%%%%%%%%%%%
    % APPS
    %%%%%%%%%%%%%%%%%%%%%%%%%%
    %
    
      \rowcolor{gray!50}  
    A1 & Telehealth &  \leftrule{\yes}&\yes&\no&
    \leftrule{\yes}&\yes & \yes&\yes &\yes&\yes &
    \leftrule{\no}&\no&\no&\yes &
     \leftrule{13} & VD &
    \leftrule{4}& 6 & 7 & 4 & 2 & 23 \\

   A2 & Telehealth & \leftrule{\no}&\no&\yes&
    \leftrule{\yes}&\yes &\yes&\yes &\yes&\no &
    \leftrule{$\circ$ }&\no&\no&\yes &
    \leftrule{13.2} & VD &
   \leftrule{2}& 5 & 7 & 4 & 2 & 20 \\
  
\rowcolor{gray!50}  
   A3 & Senior care \& Caregiver Support &   \leftrule{\no}&\no&\yes&
    \leftrule{\yes}&\no & \yes&\yes &\yes&\yes &
    \leftrule{$\circ$ }&$\circ$ &\no&\no &
    \leftrule{10.9} & FD &
    \leftrule{2}& 5 & 7 & 4 & 1 & 19 \\

 A4 &  Senior care \& Caregiver Support &
    \leftrule{\no}&\no&\yes&
    \leftrule{\yes}&\no & \yes&\yes &\no&\no &
    \leftrule{$\circ$ }&$\circ$ &\no&\yes &
    \leftrule{14.2} & P &
    \leftrule{2}& 4 & 4 & 3 & 2 & 15 \\

      \rowcolor{gray!50}
    A5 & Senior care \& Caregiver Support &  \leftrule{\no}&\no&\yes&
    \leftrule{\yes}&\no &\no&\no &\yes&\no &
    \leftrule{$\circ$ }&\no&\no &\yes &
    \leftrule{13} & VD &
    \leftrule{2}& 3 & 6 & 4 & 2 & 17 \\
    
   A6 & Senior care \& Caregiver Support &  \leftrule{\yes}&\no&\no&
    \leftrule{\no}&\yes &\yes&\yes &\no&\no &
    \leftrule{\no}&\no&\no&\yes &
    \leftrule{11.5} & D &
    \leftrule{3}& 5 & 8 & 2 & 2 & 20 \\
      \rowcolor{gray!50}
      A7 & Senior care \& Caregiver Support &  \leftrule{\no}&\no&\yes&
    \leftrule{\yes}&\yes &\yes&\yes &\yes&\no &
    \leftrule{\no}&$\circ$ &\no&\yes &
    \leftrule{12.4} & D &
    \leftrule{2}& 5 & 7 & 4 & 2 & 20 \\

    A8 & Senior care \& Caregiver Support &  \leftrule{\yes}&\no &\no &
    \leftrule{\no}&\yes &\yes&\yes &\no&\no &
    \leftrule{\no}&$\circ$ &\no&\yes &
    \leftrule{11.6} & D &
    \leftrule{3}& 5 & 7 & 2 & 2 & 19 \\

\rowcolor{gray!50}
 A9 & Senior care \& Caregiver Support &  \leftrule{\yes}&\no&\no&
    \leftrule{\yes}&\no &\yes&\yes &\yes&\no &
    \leftrule{\no}&\no&\no&\yes &
    \leftrule{13.4} & VD &
   \leftrule{3}& 4 & 7 & 4 & 2 & 20 \\

   A10 & Eldercare \& Well-being Support
 &   \leftrule{\no}&\no&\yes&
    \leftrule{\yes}&\yes &\yes&\yes &\yes&\yes &
    \leftrule{\no}&\no&\no&\yes &
    \leftrule{12.8} & VD &
    \leftrule{2}& 6 & 7 & 4 & 2 & 21 \\
 
          \rowcolor{gray!50}
    A11 & Eldercare \& Well-being Support & \leftrule{\no}&\no&\yes&
    \leftrule{\yes}&\yes &\yes&\yes &\yes&\yes &
    \leftrule{\no}&\no&\no&\yes &
    \leftrule{11.2} & FD &
    \leftrule{2}& 6 & 9 & 4 & 2 & 23 \\
 A12 &  Eldercare \& Well-being Support &
\leftrule{\no}&\no&\yes&
    \leftrule{\yes}&\yes &\yes&\yes &\yes&\no &
    \leftrule{$\circ$}&\no&\no&\yes &
    \leftrule{14.4} & P &
    \leftrule{2}& 5 & 5 & 4 & 2 & 18 \\

\rowcolor{gray!50}
         A13 & Eldercare \& Well-being Support &  \leftrule{\no}&\no&\no&
    \leftrule{\yes}&\no &\no&\yes &\no&\no &
    \leftrule{$\circ$ }&\no&\no&\yes &
    \leftrule{10.9} & FD &
    \leftrule{1}& 3 & 8 & 3 & 2 & 17 \\

   A14 & Eldercare \& Well-being Support & \leftrule{\no}&\no&\yes&
    \leftrule{\yes}&\yes &\yes&\yes &\no&\no &
    \leftrule{$\circ$}&\no&\no&\yes &
    \leftrule{13.2} & VD &
    \leftrule{2}& 5 & 6 & 3 & 2 & 18 \\

      \rowcolor{gray!50}
    A15 &   Health Monitoring \& Safety &  \leftrule{\no}&\no&\yes&
    \leftrule{\yes}&\yes &\yes&\yes &\yes&\no &
    \leftrule{\no}&\no&\no&\yes &
    \leftrule{11.5} & D &
    \leftrule{2}& 5 & 8 & 4 & 2 & 21 \\

   A16 & Health Monitoring \& Safety  &  \leftrule{\no}&\yes&\no&
     \leftrule{\yes}&\yes &\yes&\yes &\no&\no &
    \leftrule{\no}&\no&\no&\yes &
    \leftrule{13} & VD &
    \leftrule{3}& 5 & 7 & 3 & 2 & 20 \\
          \rowcolor{gray!50}
    A17 & Health Monitoring \& Safety  & \leftrule{\yes}&\no&\no&
    \leftrule{\yes}&\no & \no&\yes &\yes&\no &
    \leftrule{$\circ$ }&\no&\no&\yes &
    \leftrule{12} & D &
    \leftrule{3}& 3 & 7 & 4 & 2 & 19 \\

    A18 &   Health Monitoring \& Safety  &  \leftrule{\no}&\yes&\no&
    \leftrule{\yes}&\yes &\yes&\yes &\no&\no &
    \leftrule{\no}&\no&\no&\yes &
    \leftrule{9.2} & SD &
    \leftrule{3}& 5 & 11 & 3 & 2 & 24 \\
    
\rowcolor{gray!50}
          A19 & Health Monitoring \& Safety  &  \leftrule{\no}&\no&\yes&
    \leftrule{\yes}&\no &\yes&\yes &\yes&\no &
    \leftrule{$\circ$ }&$\circ$ &\no&\yes &
    \leftrule{11.8} & D &
    \leftrule{2}& 4 & 6 & 4 & 2 & 18 \\

   A20 & Health monitoring \& Safety &  \leftrule{\yes}&\no& \no&
    \leftrule{\yes}&\no &\yes&\yes &\yes&\yes &
    \leftrule{\no}&\no&\no&\yes &
    \leftrule{11.6} & D &
    \leftrule{3}& 5 & 8 & 4 & 2 & 22 \\

  \rowcolor{gray!50}
    A21 & Healthcare Services &  \leftrule{\yes}&\no&\no&
    \leftrule{\yes}&\yes &\yes&\yes &\no&\no &
    \leftrule{\no}&\no&\no&\yes &
    \leftrule{12.9} & VD &
    \leftrule{3}& 5 & 7 & 3 & 2 & 20 \\

   A22 & Healthcare Services & \leftrule{\no}&\no&\yes&
    \leftrule{\no}&\no &\yes&\yes &\no&\no &
    \leftrule{$\circ$ }&$\circ$ &\no&\yes &
    \leftrule{13.3} & VD &
   \leftrule{2}& 4 & 5 & 2 & 2 & 15 \\

          \rowcolor{gray!50}
   A23 & Healthcare Services  & \leftrule{\no}&\yes &\no &
    \leftrule{\yes}&\yes &\yes&\yes &\yes&\yes &
    \leftrule{\no}&\no&\no&\yes &
    \leftrule{10.6} & FD &
    \leftrule{3}& 6 & 9 & 4 & 2 & 24 \\

    A24 & Fitness Support  &  \leftrule{\no}&\no&\no&
    \leftrule{\no}&\no &\no&\no &\no&\no &
    \leftrule{\no}&\no&\no&\no &
    \leftrule{-} & - &
    \leftrule{0} & 0 & 0 & 0 & 0 & 0 \\

      \rowcolor{gray!50}
    A25 & Fitness Support &  \leftrule{\no}&\no&\no&
    \leftrule{\yes}&\no &\no&\yes &\no&\no &
    \leftrule{$\circ$ }&$\circ$ &\no&\yes &
    \leftrule{9.1} & SD &
    \leftrule{1}& 3 & 9 & 3 & 2 & 18 \\

 A26 & Fitness Support  &  \leftrule{\no}&\no&\no&
    \leftrule{\yes}&\no &\no&\yes &\no&\no &
    \leftrule{\no}&$\circ$ &\no&\yes &
    \leftrule{9.6} & SWD &
    \leftrule{1}& 3 & 9 & 3 & 2 & 18 \\
    
\rowcolor{gray!50}
   A27 & Fitness Support &  \leftrule{\no}&\yes&\no&
     \leftrule{\yes}&\yes &\yes&\yes &\yes&\no &
    \leftrule{$\circ$ }&\no&\no&\yes &
    \leftrule{12.1} & D &
    \leftrule{3}& 5 & 7 & 4 & 2 & 21 \\

    A28 &   Fitness Support &  \leftrule{\no}&\no&\no&
    \leftrule{\yes}&\yes &\yes&\yes &\yes&\no &
    \leftrule{$\circ$ }&$\circ$ &\no&\yes &
    \leftrule{11.4} & FD &
    \leftrule{1}& 5 & 7 & 4 & 2 & 19 \\

    \bottomrule
    \end{tabular}
 
    \label{tab2}
\end{table*}

\subsection{Common Gaps in Privacy Policies}

This demographic frequently has unique challenges understanding complicated terminologies and needs more precise, straightforward explanations of data handling practices. Consequently, this lack of clarity leaves them uncertain about how well their sensitive health data is protected.
Another key gap is the lack of accommodations for accessibility in any of the reviewed privacy policies which is a significant concern, considering how important accessibility is for older adults managing age-related conditions. 

% To make privacy policies more inclusive and efficient, they must be improved with language that is easy to understand and that makes special adjustments for older adults. Additionally, all apps but one failed to mention data sharing principles.

% Our assessment shows significant gaps in the adherence to privacy policies and the application of fundamental privacy principles in healthcare apps intended for older adults. The need for more privacy regulations and control in this industry is highlighted by the inconsistent adherence to regulations and the absence of transparency and accessibility considerations. It is important to enhance the clarity, accuracy, and accessibility of privacy policies to guarantee that older adults can make informed decisions regarding the privacy and security of their health information.

\subsection{Comparative Analysis of Privacy Practices }
Our review of privacy policies for the healthcare apps revealed significant differences in compliance and the implementation of privacy principles. 
% This comparative analysis highlights differences among app categories and identifies patterns and areas that require improvement.

\begin{itemize}
\item\textbf{Telehealth Apps:} Apps A1 and A2 that allow remote medical consultations should comply with privacy standards due to the sensitive nature of their data. However, our analysis suggests that only A1 explicitly referred to HIPAA and GDPR, whereas A2 broadly referenced to data privacy standards. User consent, data usage and data encryption policies were described in detail by both apps.  
However, the difference between A1 and A2's regulatory referencing indicates an opportunity to further standardize privacy policy within this category of healthcare apps.

\item \textbf{Senior Care and Caregiver Support Apps:} This category's apps displayed a mixed landscape of privacy practices and regulatory compliance. Some apps specifically referred to HIPAA $43\% $(n=$3$ out of $7$) and thoroughly explained their consent requirement indicating a commitment to user autonomy. Others like A3 and A4 made vague commitments about data security without outlining the safeguards in place which can lead to concerns about the level of protection of sensitive data. The need for better and more standardized privacy processes within this category is highlighted by the inconsistency.

\item \textbf{Eldercare and Well-being Support Apps:} All apps in this category mentioned various international data protection regulations beyond HIPAA and GDPR, except for A13, which did not reference any regulations. Generally, these apps provided detailed privacy policies addressing data minimization, user consent, and somehow access control. However, the depth of these regulations differed, with some apps providing detailed procedures and others using ambiguous language. This variability in the category implies that although we generally recognise privacy is important for health and eldercare apps, currently, there lacks a clear consistent implementation of good practice.

\item \textbf{Health Monitoring and Safety Apps:} These apps exhibited good compliance, with $4$ out of $6$ referencing HIPAA and GDPR. This compliance rate suggests that the apps recognize the necessity of adhering to data regulations due to the medical nature of their data collection during monitoring. It is laudable that apps like A15, A18, and A20 prioritized user consent and data minimization. However, there were obvious gaps in a few key areas. For example, A17 and A19 lacked detailed information on their real-time monitoring and data encryption practices, potentially leaving older adults uncertain about the security of their continuously collected health data. 
% What appeared more alarming was only a single app in this category mentioned its breach protocol.
The gap between strong regulatory awareness and inconsistent implementation stresses a need for broader approaches to privacy in health monitoring apps.

\item \textbf{Healthcare Services App:} A21 and A23 mentioned compliance with HIPAA and GDPR respectively. A23 provided detailed key principles, while A21 focused more on encryption practices and user consent mechanisms but lacked specifics on data retention periods and breach notification procedures. A22, which referenced an alternate privacy policy, failed to address data minimization and encryption protocols. The need for more precise standards on how various healthcare service types should handle data privacy is highlighted by this varied approach to regulatory compliance.

\item \textbf{Fitness Support Apps:} A27 was the only app in this category to refer to GDPR while the other four apps made no mention of any privacy policies. The lack of HIPAA references may suggest that the developers of fitness apps do not consider the data they collect to be protected health information. To guarantee the complete protection of users' health-related information, this regulatory approach may need to be reconsidered given that fitness data can indicate or reveal conditions, especially in older adults.

\end{itemize}

\subsection{Application of PRAF to Healthcare Apps}
% \subsection{Privacy Risk Assessment}

\begin{enumerate}
\item \textbf{Regulatory Compliance:}

We assessed each app's compliance with GDPR, HIPAA, and other relevant regulations. Out of the $28$ apps, $5$ apps $(17.9\%)$ claimed to be GDPR compliant and $7$ apps $(25\%)$ specifically stated HIPAA compliance. Additionally, $12$ apps $(42.9\%)$ made mention of other privacy regulations such as the California Consumer Privacy Act (CCPA), and various international data protection laws. It is interesting to note that $4$ apps $(14.3\%)$ raised questions about their commitment to standardized data protection because they did not mention adherence to any specific privacy regulations. Apps that comply with both HIPAA and GDPR received 4 points, those complying with either HIPAA or GDPR were assigned 3 points, compliance with other regulations received 2, and non-compliant apps received 1. HIPAA and GDPR received higher scores due to the extensive and significant characteristics of these standards. They both are esteemed benchmarks in data privacy for healthcare and general data protection. 

% Apps complying with other regulations received low scores, although significant, but not in the same degree of detail or rigorous enforcement as HIPAA and GDPR. 

\item \textbf{Data Security:}
This emerged as an essential component in our framework, with a mean score of $4.5$ out of $6$ (SD = $1.3$). Data encryption, access control, and breach notification protocols were the three main criteria we used to evaluate this component. Each received a score of $2$ points for implementation and $1$ point for non-implementation. About $57\%$ of apps (16 out of 28) mentioned encryption measures. For instance, A7 stated, \lq\lq Any payment transactions will be encrypted using SSL.\rq\rq~ Access control was addressed by $79\%$ of apps (22 out of 28), indicating a focus on user authentication. However, only $21\%$ of apps (6 out of 28) offered comprehensive plans for alerting users to potential data breaches, indicating a significant gap in incident response readiness. 

% This finding emphasizes the necessity of more comprehensive security measures that cover both preventive and reactive methods.

\item \textbf{Usability and Accessibility:}

These were critical components in our Privacy Risk Assessment Framework, with a mean score of $7$ out of $12$ (SD = $2$). We evaluated this based on readability (using the SMOG index) ~\cite{readabilityformulasReadabilityScoring}, clarity of policies, and the presence of accessibility features. We scored readability on a scale of 1-6 (Professional=1, Very Difficult=2, Difficult=3, Fairly Difficult=4, Somewhat Difficult=5, Slightly Difficult=6). Policy clarity was scored 2 points for clear, 1 for vague or ambiguous language, and accessibility scored 2 points for present, and 1 for absent. Despite the importance of these features for older adults, none of the apps included specific accessibility options, which are crucial for this demographic.

\item \textbf{Data Minimization and Retention:}

These practices were evaluated as part of our Framework, with a mean score of $3.4$ out of $4$ (SD = $1$). Each components were assigned 2 points for implementation and 1 point for non-implementation. While all apps claimed to collect only necessary data, only 86\% (24 out of 28) provided specific details on their data minimization practices. For example, A3 stated, \lq\lq We limit the collection of personal information to what you choose to submit through the use of our services\rq\rq~, but some apps like A6 and A22 did not mention data minimization commitments. 
Regarding data retention, 57\% (16 out of 28) of the apps specified retention periods, with A19's policy indicating a 5-year retention time. Concerns regarding possible over-collection and indefinite storage of sensitive health information are highlighted by the lack of specificity in data collection extent and retention periods in some of the apps.

\item \textbf{Third Party Data Sharing:}
This emerged as a critical component in our Framework, with a mean score of $1.9$ out of $2$ (SD = $0.4$). We evaluated this based on transparency of sharing practices, disclosure of third parties, and user control.  This was assigned a score of 2 points for implementation and 1 point for non-implementation. Nearly all apps (26 out of 28) included third party sharing details, outlining who, when, and why data might be shared. The emphasis on third-party sharing practices highlights its importance in the overall privacy protection for older adults using healthcare apps.

\end{enumerate}
% The overall privacy risk scores for each app are in Table~\ref{tab2}.
% The variations in regulatory compliance among these apps demonstrate the fragmented nature of privacy protection in the healthcare app sector for older adults. 

% These variations highlights the difficulties older adults encounter while navigating the complex landscape of digital health privacy and highlights the need for more standardized and strict privacy regulations in this rapidly expanding field.

%-------------------------------------------------------------------------------
\section{Discussion and Implications}
%-------------------------------------------------------------------------------
% We evaluated the comprehensibility and transparency of the privacy policies of our healthcare apps (\textit{RQ1}). Many privacy policies were difficult to understand due to unclear language. Transparency varied significantly, with some apps providing detailed explanations while others remained vague about their practices. The second research question addressed regulatory compliance and inconsistencies in the implementation of privacy standards in healthcare apps for older adults (\textit{RQ2}). The third research question examines privacy risks across multiple dimensions (\textit{RQ3}). 

\subsection{Comprehensibility of Privacy Policies for Older Adults -- RQ1}
\begin{figure*}[htbp!]
\centering
\includegraphics[width=0.75\linewidth]{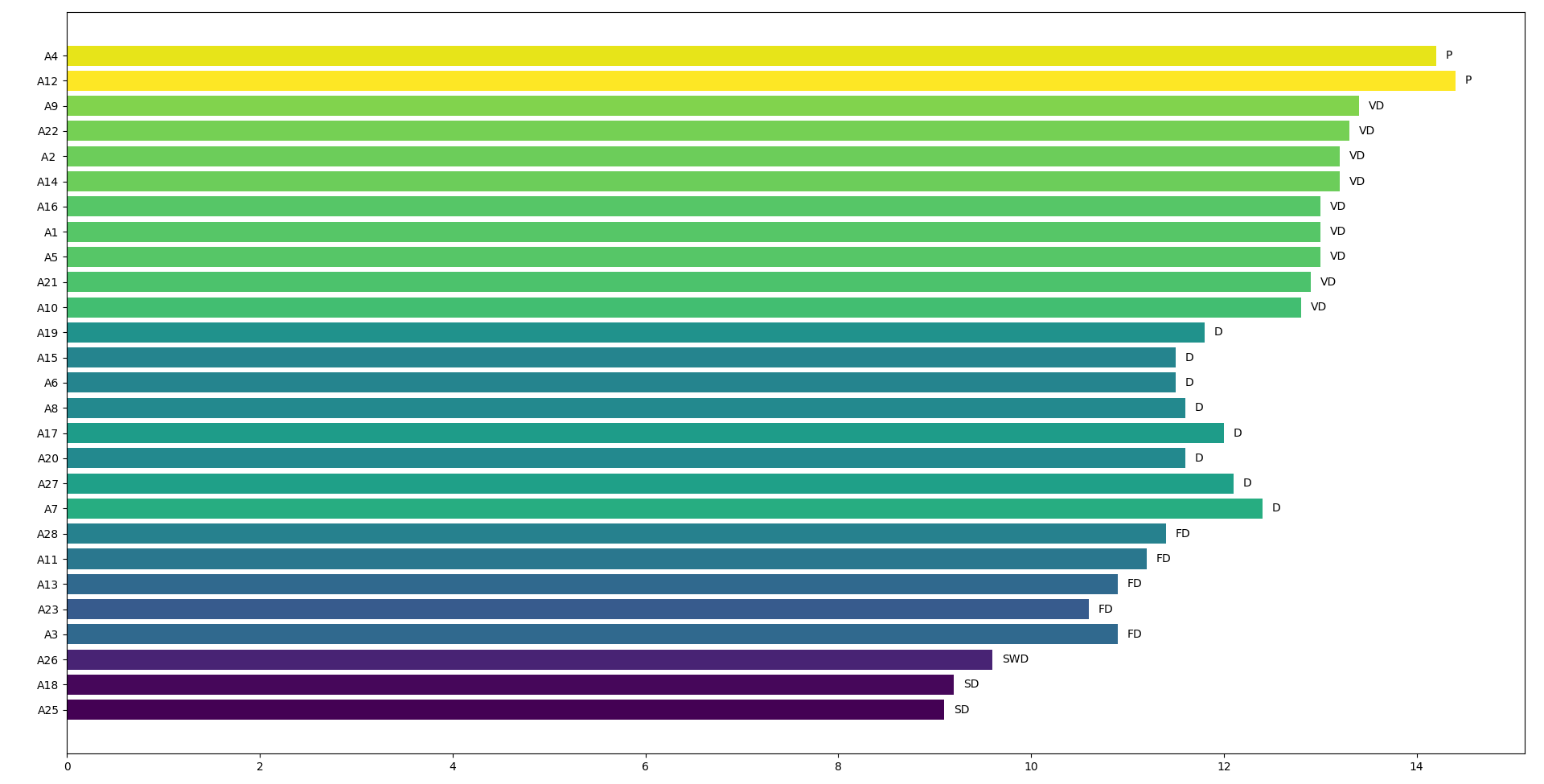}
\caption{Readability of Healthcare Apps using the Simple Measure of Gobbledygook (SMOG) Index}\label{smog}
\label{process}
\end{figure*}

In our analysis, we observed significant challenges with the readability and clarity of privacy policies. These issues raise concerns about the ability of this population to make decisions regarding their health data. Readability was a major concern since about half of the apps under review used complex terminology. The use of choice of phrasing presents a significant barrier to understanding, particularly for older adults whose health and digital literacy vary. To quantify this issue, we evaluated all the collected privacy policies using the SMOG (Simple Measure of Gobbledygook) Index readability test. The SMOG test measures the readability of complex data to make sure the content is clear and understandable to older adults~\cite{Faroog_SMOG_2020, Hedman_smog_2008, fahimuddin2019readability}. Using the Readability Formulas calculator ~\cite{readabilityformulasReadabilityScoring}, we discovered that the average grade level was roughly $12$, which indicates a difficult reading level and the readability score is detailed in Table~\ref{tab2} and graphically represented in Figure~\ref{smog}. This highlights the necessity for privacy regulations to use accessible language, especially for apps targeted at older adults who can encounter age-related cognitive challenges that impair their ability to understand text. 
% Our analysis also identified some lack of transparency in data collection and usage practices among the studied apps.
% Some apps offered generic statements, but others gave detailed descriptions of the data collected and how it is used. For instance, statements such as "we collect data to improve our services" fall short of providing sufficient disclosure since they don't address the specific categories of data that are collected or how they are used. 
Older adults who might already be cautious of digital privacy issues need clear and detailed information on what personal data is being gathered, how long it will be retained, what security measures are in place, and how their data is being processed.

% Informing users of their rights to personal data, such as the right to access, modify, or delete it, is another aspect of transparency.

\subsection{Regulatory Compliance and Gaps in Privacy Protections -- RQ2}

Our study identified several key principles and protections commonly missing from these apps' privacy policies. only $4$ apps failed to indicate the precise extent of data collected, indicating that data minimization was well addressed. However, only $57\%$ ($16$ apps) described their data retention periods, and even fewer ($21\%$) outlined the steps involved in notifying users of a data breach. This could mean that many users are not aware of how they would be notified in the event of a data breach. There were only two apps that did not include user consent in their privacy rules, making it the principle that was most often addressed.
While HIPAA and GDPR provide important privacy protections, their effectiveness in regulating healthcare apps for older adults seems to be limited. Many healthcare apps are operating in a regulatory gray area due to HIPAA's narrow focus on covered entities. Our finding that only $25\%$ of apps specifically reference HIPAA implies that a significant portion may fall outside its jurisdiction. The global nature of the app market complicates GDPR enforcement, especially for developers who are not situated in the EU. This could account for our study's lower GDPR compliance number $(17.9\%)$. HIPAA and GDPR fall short of addressing the unique challenges presented by mobile health technologies, such as the ongoing collection of data via wearables or IoT~\cite{toutsop2021exploring,gopavaram2021iot,hadan2019making,tazi2023accessibility,saka2023safeguarding}, which are becoming more relevant for older adults' care. There are significant gaps in basic privacy principles and protections, which call for a reassessment and possible expansion of existing regulatory systems to guarantee comprehensive privacy protections for older adults.

\subsection{Policy and Practice Implications -- RQ3}
Our findings of varying degrees of regulatory compliance highlight the need for a more uniform and comprehensive regulatory framework. One of the most pressing implications of our study is the need to strengthen regulatory frameworks like HIPAA and GDPR. While these bodies provide a solid foundation for data protection, our findings indicate that compliance is inconsistent across different types of healthcare apps. There is a clear need for more specific guidelines tailored to the unique contexts of various app categories, particularly those targeting older adults. Regulatory bodies should consider developing tailored guidelines and increasing regulatory checks and balances for adequate enforcement of privacy policies by designers, this will in turn ensure that all healthcare apps comply with established privacy standards. Beyond regulatory change, our study suggests improvement of the healthcare app industry by developing standardized privacy policy templates that emphasize plain language, clear explanations of data handling, and user rights on data. Also, HIPAA and GDPR should be amended to specifically include mobile health technology and continuous data collection to handle the evolving field of digital health. To give older adults a reliable indication of an app's credibility, we can establish a standardized Privacy Seal or Stamp certification for those that adhere to international privacy standards such as nutrition labels~\cite{Kelley_2009_NutritionLabel}. Our findings highlight the pressing need for industry and regulatory changes to improve data privacy for older adults using healthcare apps. We can make the environment more reliable and secure for older adults by strengthening regulatory frameworks, harmonizing industrial practices, and concentrating on efficient regulatory implementations
% Harmonizing international standards for data privacy is crucial, especially considering the worldwide reach of app markets. Establishing a global collaborative body to create cross-border enforcement procedures for privacy is what we advise doing to guarantee uniform protection across borders. 

\subsection{Implications of PRAF}
The framework we introduced offers a thorough assessment of the privacy policies of healthcare apps with a particular emphasis on how well-suited they are for older adults. We can identify major privacy concerns and provide helpful recommendations for enhancing these apps' privacy policies by analyzing the privacy risk scores across critical components.
The PRAF's findings show that the evaluated healthcare apps have varying levels of compliance, with some critical areas in need of major improvement. The average regulatory compliance score was $2.2$ out of $4$, meaning that while some apps comply with regulations like HIPAA and GDPR, many do not. The range of scores from $0$ to $4$ indicates that different apps apply the regulations in different ways, which could make some older adults more susceptible to privacy breaches.
A relative strength that stood out was data security, with an average score of $4.5$ out of $6$. This suggests that most of the apps talked about fundamental security features like data encryption and access control. However, inconsistent practices remain a concern, especially about breach notifications. The standard deviation indicates a moderate degree of implementation, with some apps like A1 and A23 scoring high at $6$ and others like A25 scoring much lower at $3$. Due to the sensitivity of health information, older adults may not be well prepared to handle the consequences resulting from data breaches, thus minor security gaps can be highly harmful.

The mean score for usability and accessibility was $7$ out of $12$, indicating that while some apps meet the needs of older adults, some do not, which calls for more consistent standards to handle physical and cognitive challenges. This is especially concerning because accidental privacy breaches may occur if older adults find it difficult to understand or use privacy settings.
Data Minimization and Retention practices with a mean score of $3.4$ out of $4$ show relatively consistent implementation across apps. While this is positive, the fact that not all apps receive the highest score implies that some may still be gathering or retaining more data than is required, which might put older adults at greater risk of privacy violations. Finally, Third-Party Data Sharing with a mean score of $1.9$ out of $2$ appear to be the most consistently implemented component with just two apps failing to include this in their privacy policies.
The average overall risk score was $18$ out of a possible $28$, with scores ranging from $15$ (A4) to $24$ (A18 and A23). More than half of the apps have scores ranging from $18$ to $22$ which suggests that they have moderate privacy risks. This finding is very significant given the targeted group and the sensitivity of health-related data. According to our framework analysis, some apps often perform poorly in certain areas but excel in others. This emphasizes the necessity of a comprehensive privacy protection strategy considering older adults' requirements and challenges. App developers and policymakers could help create a safer and more secure digital health environment that better serves this demographic by addressing the privacy risks highlighted in our study.

Based on our findings, we suggest that targeted regulations be put in place to better meet the privacy demands of older adults who use healthcare apps. It is recommended that policymakers create standards that take into account the intersection between aging, technology, and health data privacy. It is the responsibility of developers to enhance usability and accessibility by designing user-friendly, age-appropriate interfaces with features like larger text options and simpler language. While maintaining strong data security is essential, usability should not be compromised; innovative ideas are required to strike a balance between the two. 
% Furthermore, it is critical that data practices be more transparent and that older adults receive clear communication about the collection, use, and sharing of their data. 
PRAF can be a useful tool for continual assessment in identifying and reducing privacy risks for older adults.

%-------------------------------------------------------------------------------
\section{Limitations and Future Work}
%-------------------------------------------------------------------------------

While our sample size is small, it is appropriate given our specific focus on healthcare apps for older adults. The lack of a dedicated \lq\lq older adults\rq\rq~ category in app stores and the absence of pre-existing datasets limited our ability to expand the sample further. It is important to note that our study represents a snapshot of the current landscape, which is subject to change due to the dynamic nature of the mobile app market. Mergers, new developments, or discontinuations of apps could alter this landscape. Additionally, our search results may be influenced by country-specific app store outputs, potentially omitting some apps due to geographic restrictions.
We recommend that future research expand on assessing the long-term impact of privacy practices to gain more understanding of how well current measures protect older adults' data over time.

% Additionally, more research should focus on how emerging technologies like artificial intelligence and blockchain can provide stronger data protection for older adults. 
% Our analysis relies on a degree of subjective interpretation of privacy policy language because there is no standard tool for rating the level of privacy protection. We addressed this by using a consensus approach, a technique used in related review~\cite{ROSENFELD_dementia_2017}, to balance variations in subjective interpretation. 

%-------------------------------------------------------------------------------
\section{Conclusion}
%-------------------------------------------------------------------------------
This study provides an assessment of the privacy concerns of healthcare apps designed for older adults using a multi-dimensional Privacy Risk Assessment Framework (PRAF). Our finding shows significant gaps in privacy protection, with many apps falling short of older adults' unique requirements. A major barrier to understanding privacy policies is the average readability level of grade 12 from our analysis. Additionally, just $25\%$ of apps specifically stated HIPAA compliance, and only $18\%$ cited GDPR, which is a concern for international privacy standards. The gap between real practices and regulatory standards is further demonstrated by the fact that $79\%$ of apps lack breach notification procedures. These findings highlight the critical need for a paradigm shift in the way that privacy is conceived of and applied in healthcare apps for older adults, emphasizing the importance of focused regulations and clearer guidelines that address the intersection of aging, technology, and health data privacy. To better protect older adults, developers must address the unique privacy requirements of this demographic by improving healthcare apps' transparency, security, and user control in addition to complying with regulatory standards.

\section{Acknowledgement}
We would also like to thank our participants for this study and acknowledge the Inclusive Security and Privacy-focused Innovative Research in Information Technology (InSPIRIT) Lab at the University of Denver for supporting this work. Any opinions, findings, conclusions, or recommendations expressed in this material are solely those of the authors.
\bibliographystyle{plain}
\bibliography{BuildSecIoT}

\end{document}